# How might humans terraform Mars?
## A rapid explanation through one example of how terraforming might occur.

## An Introduction to Mars Terraforming
2025 Summary provided by Pioneer Labs
Devon Stork, Erika DeBenedictis

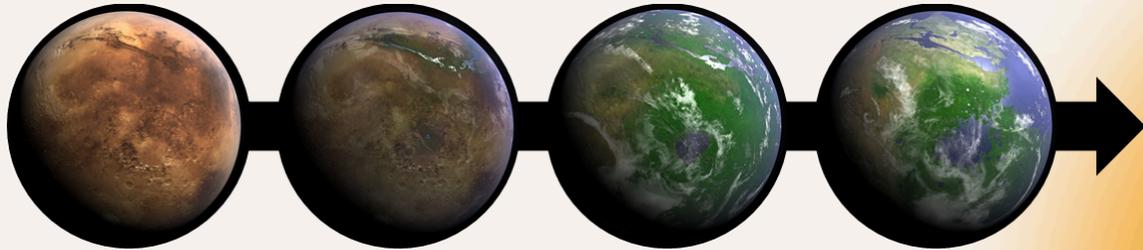

Welcome to the 2025 Green Mars workshop! **This document presents one story of how terraforming Mars might occur as a way to rapidly bring attendees up to speed.** It incorporates results from last year's workshop and the subsequent Nature Astronomy perspective, and builds on research by many groups over the past 50 years.

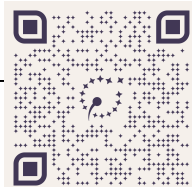
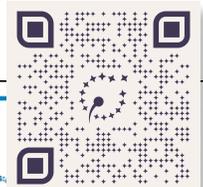

Today, Mars terraforming is a vision, not a detailed plan. As McKay wrote in 1991, "The complexity of the problem of terraforming a planet does not allow us to draw detailed conclusions. The problem is similar in scope to understanding the Earth's biosphere and biogeochemical cycles."

Throughout this document, in addition to the example terraforming story, we will also call out:

| Key unknowns and research priorities: | Alternatives: | Immediate benefits: |
|---|---|---|
| Much remains to be studied. Throughout, we will highlight key scientific unknowns and critical areas for further research. We will also note **Potential Showstoppers** - problems that might make this particular terraforming story infeasible. | For the purposes of explanation, we highlight just one story for how Mars might be terraformed. There are many other options, and we will describe these for consideration. | Green Mars research has the potential to advance science and technology in many other areas, both in space exploration and terrestrially. |

# There are many possible visions for Mars's future
'Terraforming' means different things to different people.

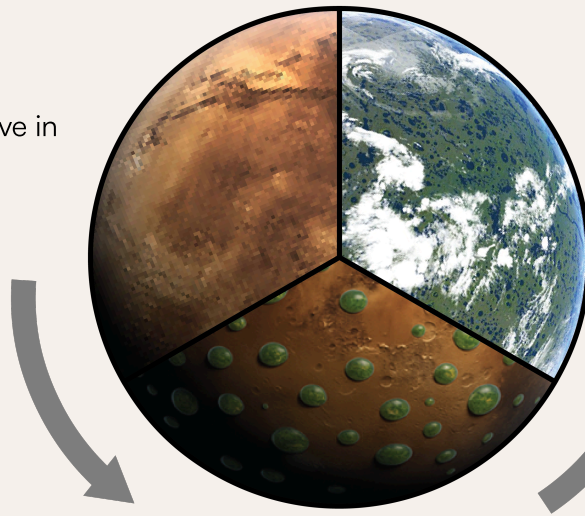

**Local Outposts**
Preservation of modern Mars surface, humans live in habitats.

**Global Terraforming**
Humans can walk around outside without a pressure suit.

**Regional Terraforming**
Regional habitats, ecosystems & cities are enabled by large structures or localized warming. Sometimes called "paraterraforming."

Technical feasibility is not the only consideration for terraforming. Cultural priorities and ethical boundaries also shape our vision for the desired endpoint. Creating sustainable habitats and ecosystems beyond Earth[1] can be approached in (at least) three ways. These visions are largely sequential, i.e. local terraforming would likely precede global terraforming.

## 01
### Local Outposts
Mars would be left primarily as a natural preserve, similar to the modern Antarctica. Humans would live in local habitats or underground, using Martian resources to survive.[2] This goal is the most near-term accessible.

## 02
### Regional Terraforming
Mars would host large domes or enclosed structures built from local materials and capable of sustaining humans and/or ecosystems.[3] These engineered habitats would create pockets of habitability; most of Mars would remain cold and dry. Mars-based production of the necessary materials could happen over decades.

## 03
### Global Terraforming
Mars would have a oxygen-rich atmosphere, streams, lakes and flowering plants.[4] The planet as a whole becomes habitable over the course of many centuries.

**Key unknowns and research priorities:**
The financing, governance, ethical, and legal framework for Mars terraforming remain highly uncertain. Although not the focus of this workshop, defining these dimensions is as critical as solving the technical challenges of habitat and ecosystem design.

**Alternatives:**
Free-space or asteroid settlements might similarly allow sustainable habitation beyond Earth. Both approaches require advances in micro-gravity habitation & life support, and careful conservation of life-essential volatile elements.

**Immediate benefits:**
Research and development of sustainable habitats on Mars drives advances in life support, resource use, and ecosystem engineering with benefits on Earth. Terraforming research is enriched by, and also supports, climate science, biotechnology, and closed-loop agriculture. Early human missions provide testbeds for these technologies.

# What would a globally terraformed Mars look like?

Humans would enjoy a planet-wide green biosphere and walk around outdoors without a pressure suit thanks to a thin $O_2$ atmosphere.

A thin, oxygen-rich atmosphere is breathable by humans and blocks most radiation

Stronger day-night temperature cycles than on Earth

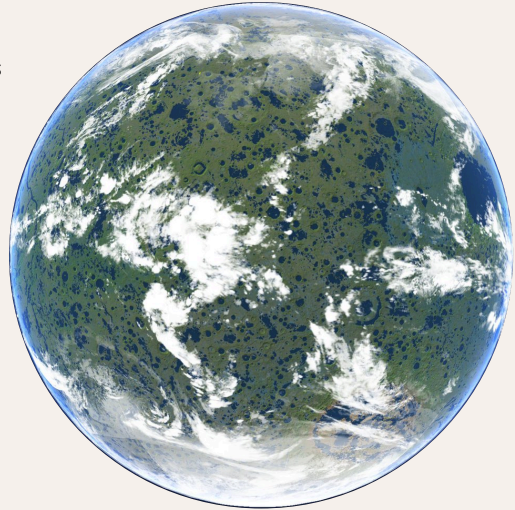

Alpine climate with 5-10% of the surface covered by lakes and seas

No global magnetic field

Photosynthetic plants

This vision of a fully terraformed Mars uses local atmosphere, water, and regolith to build a thin but breathable climate, stable ecosystems, and a habitable surface without importing materials from off-planet. This endpoint could likely be achieved over many centuries without major new spaceflight capabilities beyond the current "Starship Era."

**Atmosphere of ~150 mbar $O_2$ and minor $N_2$.**
- Created through photosynthesis, like Earth's $O_2$.
- Breathable by humans without a mask.[5]
- Below flammability limit.[6]
- Likely to block a significant fraction of ultraviolet radiation through ozone evolution.[7]
- Solar energetic particles blocked by atmospheres over 12 mbar.[8]
- Reduces galactic cosmic radiation to levels similar to ISS, or 2.4 m of regolith.[9]

**No magnetic field**
- Mars' atmospheric loss rate is $5 \times 10^7$ kg/year[10]. This can be ignored for at least 100 million years.
- Surface radiation is substantially reduced by the thin atmosphere, see notes in the left column).

**Global average temperature of -5°C**, about 20 °C colder than Earth.
- This temperature retains surface permafrost, preventing loss of water to subsurface aquifers.
- Warm days, cold nights. The thin atmosphere results in substantial day/night temperature swings.

**Key unknowns and research priorities:**
- What are the long term human health risks of a 150 mbar $O_2$ atmosphere and exposure to Galactic Cosmic Radiation? Could health risks be mitigated?
- Accurate estimates of UV protection from a 150 mbar $O_2$ atmosphere demand a photochemical model of ozone dynamics.
- Organisms on Mars could obtain nitrogen from the regolith. However, nitrogen released to the atmosphere during rotting must be recaptured, requiring study of low-pressure nitrogen fixation.
- **Potential Showstopper:** On a warmed Mars, water might become trapped as ice at cold, high-altitude locations. Additional climate modeling is needed to understand this risk.

**Alternatives:**

Warmer average temperature would be desirable, as it would result in a wetter and more metabolically active biosphere. The story above assumes that permafrost is used to seal liquid water on the surface and prevent loss to empty subsurface aquifers, if they exist. If they do not exist, the target average global temperature could be substantially increased.

# What is different about a Green Mars?
Two major changes would transform the planet. 1) warming melts ice and 2) photosynthesis converts some of the former ice into an oxygen-rich atmosphere.

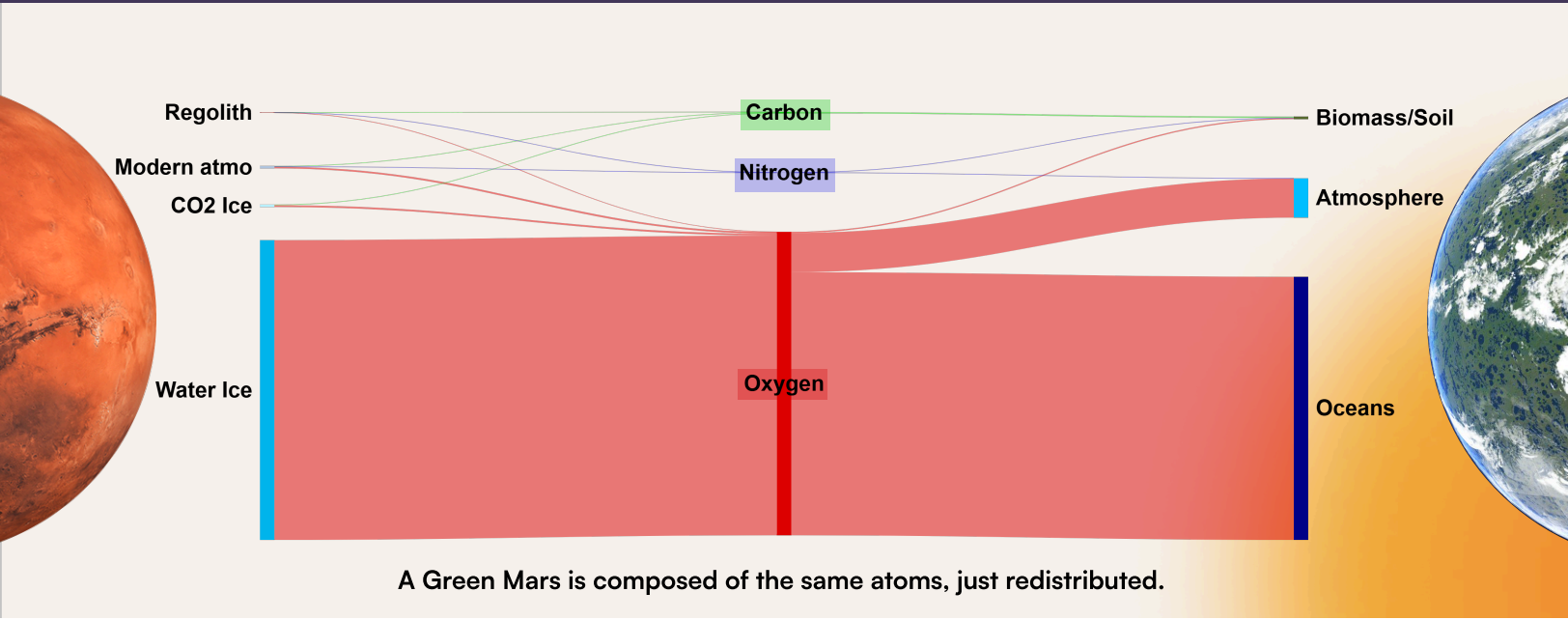

A Green Mars is composed of the same atoms, just redistributed.

For a version with citations & log-scaling, see appendix.[11]

**Existing raw materials include**

- **Regolith.** Martian regolith contains fixed nitrogen in the form of nitrate, eliminating early needs for nitrogen fixation.
- **Atmosphere.** Most of the modern $CO_2$ rich atmosphere is used as the carbon source for biomass, to sink hydrogen.
- **$CO_2$ ice cap.** Radar data indicate that the south polar $CO_2$ ice deposit mass is only ~24 teratonnes, nowhere near enough to create an atmosphere if melted, as has been proposed.[12]
- **Ice.** The above ice estimate includes only near-surface water.[13] It is likely there are also substantial quantities of subsurface water.

**Breakdown of terraformed end state**

- Biomass distributed between live biomass and dead humic matter in the soil. The upper limit is roughly 25x Earth's biomass.
- Atmosphere is 99.9% $O_2$, 0.1% Nitrogen
- The oceans contain ~2% as much water as Earth, mostly pooling in the northern lowlands.
- Regolith is supplanted by soil.

**Key unknowns and research priorities:**

- How much water (ice and liquid) exists on Mars deep underground?
- **Potential Showstopper:** Oxygen build-up in the atmosphere implies that hydrogen from water is sequestered (analogous to organic-matter burial from Earth). We do not know if Mars contains enough electron acceptors (carbon, ferric iron, sulfate) to serve as a hydrogen sink and liberate the necessary amount of oxygen from water for the desired atmosphere.

**Alternatives:**

- This possible endpoint uses atoms already on Mars. If additional resources are desired, one could direct additional volatile rich asteroids or comets towards Mars. However, this is energetically expensive and beyond the spaceflight capabilities of the current "Starship Era."
- Local terraforming approaches would do similar transformations on the smaller scale, allowing concentration of limiting resources.

# How could we create a planetary-scale biosphere?

Planet scale redistribution of materials requires a large, active biosphere to perform photosynthesis. Creating an active biosphere requires enough heat to maintain liquid water.

### Solid-state Greenhouse

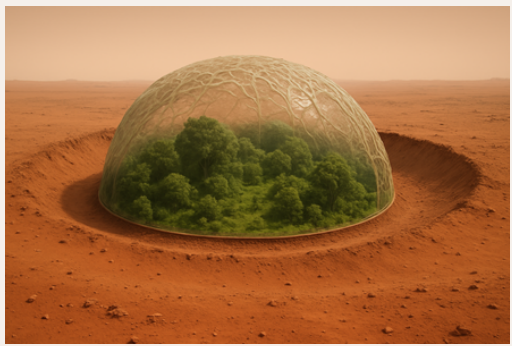

Use transparent insulators like aerogels to create local habitable zones

### Gaseous Greenhouse

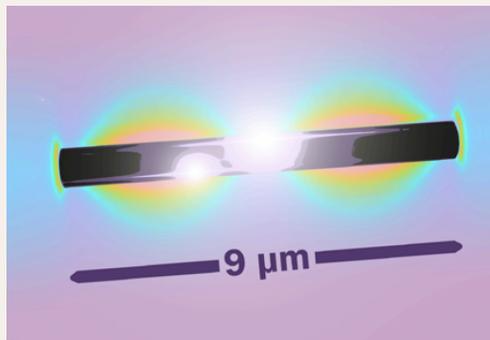

Aerosol 'glitter' that reflects IR radiation back toward the surface, creating a greenhouse effect.

Many microbes and some plants can grow at low pressure,[16] making temperature a more serious barrier to plant growth. A combination of approaches could be used to create habitats at scale that are warm enough for photosynthetic organisms. Crucially, a warmer Mars could enable liquid water even at low atmospheric pressure.

**Insulating materials[14]**

- Materials that are transparent and insulating like silica aerogels[17] and cellulose aerogels[18] can be used to create greenhouses.
- Heat will build up underneath these insulators over time.
- Presence of the insulator will buffer the day/night temperature swing, which would be extreme even on a warmer Mars.

**Heating with conductive aerosol nanoparticles[15]**

- Such particles can be used to reflect IR radiation back toward the surface.
- Very mass efficient.[19]

**Ineffective heating techniques**

- Nuclear weapons are infeasible.[20]
- There's insufficient fluorine on Mars to warm the planet with conventional greenhouse gases.[21]

Meaningful warming could be achieved quickly. The global average temperature could be increased 30°C in 30 years, enough to enable scalable plant growth outside. This plant cover would generate enough oxygen to fill a dome habitat in ~2 years, and accumulate a planetary scale atmosphere over 10,000 years.[22]

**Key unknowns and research priorities:**

- How do we make self-growing insulators with biology?
- How to manufacture greenhouse aerosols?
- How much does the loss of $CO_2$, now sequestered as biomass, hurt the greenhouse effect?
- **Potential Showstopper:** Are aerosol 'glitter' nanoparticles toxic to humans?

**Alternatives:**

Solar sails could alternatively be used to heat the planet globally or locally.[32] Preliminary calculations at last year's workshop suggest this may be similarly cost effective to the methods listed above.

**Immediate benefits:**

- Paraterraforming, or covering large fractions of the planet with habitable greenhouses, is its own desirable endpoint.
- Domes with plants would fill with oxygen in a couple years, **allowing habitation & homesteading within our lifetimes**.

# How could we grow the first organism?

Ecological succession begins with a pioneer species that is the first to repopulate a harsh environment, preparing the way for a more a complex ecosystem.

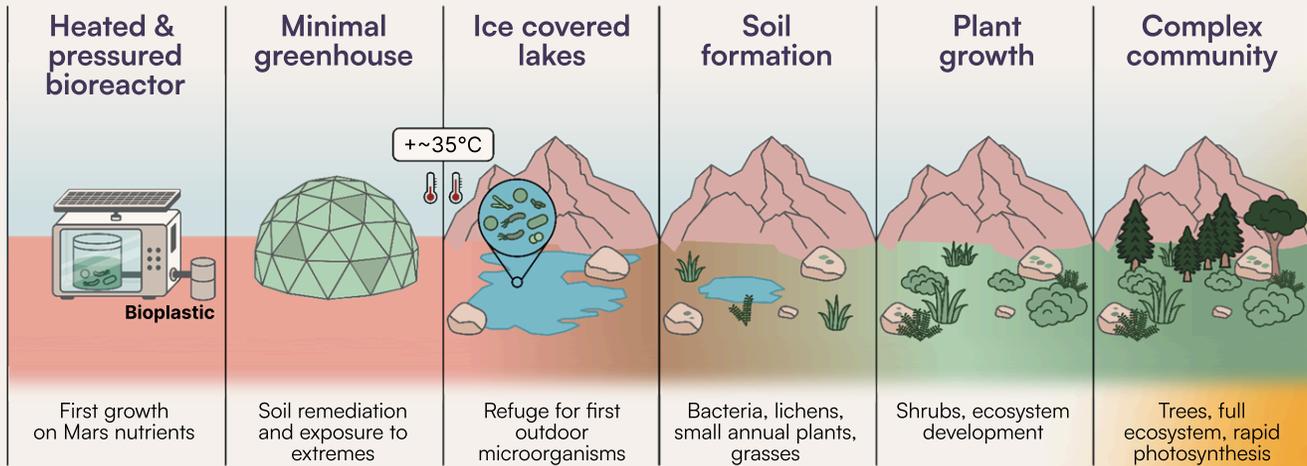

| Heated & pressured bioreactor | Minimal greenhouse | Ice covered lakes | Soil formation | Plant growth | Complex community |
|---|---|---|---|---|---|
| First growth on Mars nutrients | Soil remediation and exposure to extremes | Refuge for first outdoor microorganisms | Bacteria, lichens, small annual plants, grasses | Shrubs, ecosystem development | Trees, full ecosystem, rapid photosynthesis |

Initial steps of ecological succession would depend on artificial support, beginning with controlled bioreactors moving towards outdoor ecosystems. Once liquid water is present on the surface of Mars, Pioneer species — microbes, lichens, and hardy plants — would prepare the Martian surface for more complex life by remediating perchlorate, boosting $O_2$, and providing a source of bioavailable nutrients through soil formation.

### Pioneer Species

- Martian resources (ice, regolith, atmosphere) contain the raw nutrients for life.[23]
- However, a pioneer species needs to survive inherent Martian stresses including perchlorate and salinity.
- Before warming, controlled environments are necessary for UV & temperature protection.
- Biological processes can convert raw Martian resources into structural materials for minimal greenhouses,[24] creating an opportunity to 'bootstrap' toward greater biomass.

### Ecological succession

- Similar to Antarctic analogs, the first outdoor growth would likely occur in ice-covered lakes.[25]
- Biological insulation may be necessary to survive day-night temperature swings.
- Liquid water enables soil formation via microbial activity through organic acid and protein weathering.[26]
- Higher-order plants accomplish oxygenation through photosynthesis.

#### Key unknowns and research priorities:

- What are the concentrations and spatial distribution of soluble nitrate, phosphate, sulfate, and trace metals on Mars? Soil volatiles (especially nitrate) data come from just two sites - how generalizable are they?
- What are the engineering targets and environmental conditions of "minimal greenhouses"?
- **Is there already life on Mars?** If there is, it may change the 'should we', or create opportunities to work symbiotically with existing Martian life.[27]

#### Alternatives:

Instead of transforming the whole planet, local terraforming could advance through networks of self-expanding greenhouses. This paraterraforming is its own desirable endpoint.

#### Immediate benefits:

First human missions in isolated outposts are their own desirable endpoint. **The first human missions will be greatly aided by biology.** Soil remediation, agriculture and farming in greenhouses, scalable *in situ* productions of biomaterials for habitats are all enabled by organisms capable of growth on Mars.

# How could we de-risk critical technology?
## Flight tests will de-risk key technologies in the next decade.

### Relevant missions/payload concepts include:

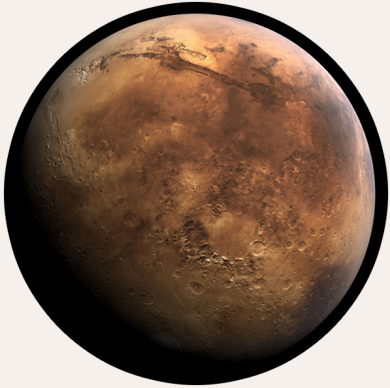
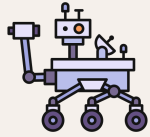

**Landers:**
- NASA-led Mars Life Explorer (Decadal-prioritized)
- Italian Space Agency plant growth experiment
- 2031 Abiotic warming + biological ISRU test mission
- Sample returns to Phobos (Japan) and Mars (China)
- Commercial Mars Payload Services

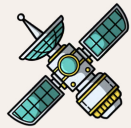

**Orbiters:**
- International Mars Ice Mapper
- Mars Telecommunications Orbiter
- Commercial Mars Payload Services
- Solar mirror experiments

Many key missions have already been proposed that would contribute to the vision of a Green Mars while also forwarding existing scientific and engineering priorities. For instance,

- The search for Life Science Analysis Group (SFL-SAG)[28] seeks extant life and to assess modern habitability and climate evolution on Mars.
- The ASI plant growth experiment[29] is testing plant growth to inform life support systems.
- 2031 Abiotic warming + biological ISRU Lander would derisk aerosol dispersal & optical properties and confirm biological growth on Martian resources.

- Non-polar ice-mapping to inform mission sites with the International Mars Ice Mapper.[30]
- Mars Telecommunications Orbiter,[31] which would include a module to test climate models.
- Solar mirror experiments[32] to de-risk this approach to local & global Mars warming.

### Key unknowns and priorities:

- Unknowns in payload profile, including mass and power needs that need to be solved.
- Availability and timeline of Starship flights for Mars transport.
- **Potential Showstopper:** Policy & regulatory control of spaceflight, especially Planetary Protection.

### Alternatives:

- Mars sample return would be an invaluable alternative to testing growth of organisms on Mars' surface. Sample return would allow thorough characterization of the chemical composition of the regolith, and could even enable growth of plants in actual regolith samples, as has been done for lunar regolith.[33]
- Field tests can de-risk many aspects of hardware design and reliability engineering. Tests in LEO or on the moon might de-risk hardware reliability during launch, but not de-risk Mars-specific climate or chemistry.

### Immediate benefits:

Flight-tested hardware for off-grid biological ISRU in harsh environments will find other applications terrestrially for manufacturing in low-resource environments like the battlefield or the third world.

# How to begin? Community-wide discussion today!
The 2025 Green Mars workshop brings together the research community to discuss research priorities, key questions, and next steps.

This year's workshop builds on momentum from last year.

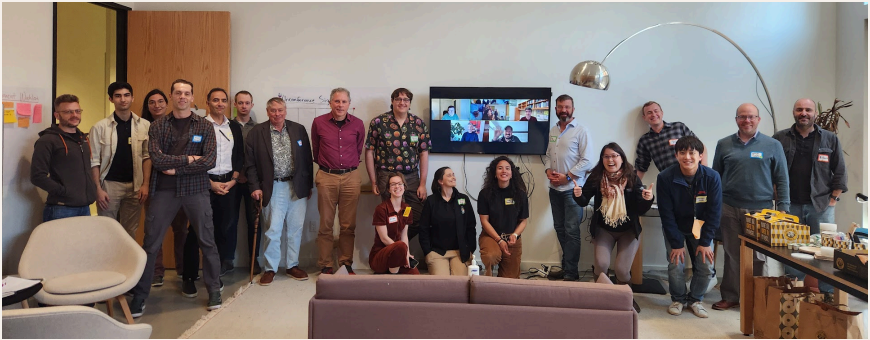

## 2024 Workshop
Last year's workshop: is terraforming Mars feasible, how could it be done, and what might change our minds?

Mars terraforming research provides fresh impetus for many existing community priorities, including climate modeling, planetary science, mission design, astrobiology, biological engineering & astronaut safety. A palette of possible priorities (whose relative importance might be discussed at this workshop) might include:

**These efforts build on previous work:**
- Escape of the martian protoatmosphere and initial water inventory, **Lindy Elkins-Tanton** 2008,
- The Mars Dust Cycle. The Atmosphere and Climate of Mars **Melinda Kahre** et al, 2017.
- Overview and Status of NASA's Planetary Protection Mission Portfolio. **Betsy Pugel** et al, 2018.
- Microbial biomanufacturing for space-exploration—what to take and when to make, **Nils Averesch** et al. 2023.
- A Biologist's Guide to Mars Dirt, **Una Nattermann** 2024.
- Sustained and comparative habitability beyond Earth, **Charles Cockell** et al. 2024
- Biomaterials for organically generated habitats beyond Earth, **Robin Wordsworth** et al. 2025.
- The Case for Mars, **Robert Zubrin**, 1996.
- Mapping Mars: Science, Imagination, and the Birth of a World, **Oliver Morton**. 2003.
- How to get to Earth from Mars: Solving the hard part first, **Casey Handmer**, 2017.
- Philanthropic Space Science: The Breakthrough Initiatives **Pete Worden** et al. 2018.
- Destination Mars: 'Innovating For Space Makes Life On Earth Better' **Tiffany Vora**, 2025.

**Terraforming research offers new approaches to key research questions:**

Climate Modeling
- Improved climate models to understand pressure & temperature cycles.
- What is the fate of water on a warmed Mars?

Habitat & Mission Design
- Can Mars-sourced biomaterials be used for crop growth or human habitats?
- What are the design parameters & construction methods for habitats & minimal greenhouses?

Biological Engineering
- What pioneer species are well suited to the extremes of Mars?
- Requirements for bioremediation & soil formation?

Planetary Science
- Improved data on biologically relevant nutrients.
- More accurate estimates of carbon dioxide and water ice reservoirs on Mars?

Astrobiology
- Firmer conclusions on extant Martian life?
- Experimentation to establish the limits of life in various stages of Mars warming & minimally supported environments?

Human Habitation
- Effects & mitigation of a 150 mbar $O_2$ atmosphere, aerosol 'glitter', and long-term Galactic Cosmic Radiation effects?

# Appendix

**1:** As described in eg. Applied Astrobiology: An Integrated Approach to the Future of Life in Space, *Wordsworth et al. 2025*

**2:** A local habitat would draw resources from Mars, eg. Towards a Biomanufactory on Mars, *Berliner et al. 2021*

**3:** Solid materials such as aerogels could passively support habitable zones, reported in Enabling Martian habitability with silica aerogel via the solid-state greenhouse effect, *Wordsworth et al. 2019.*

**4:** Global terraforming, as introduced in On Terraforming Mars, *Chris McKay. 1982.*

**5:** 150 mbar $O_2$ is above the Armstrong limit of 62 mbar, where water will boil at human body temperature, described in Human Exposure to Vacuum, *Geoffrey A. Landis, 2000*. It is also above the partial pressure of oxygen at the peak of mount Everest, which at 56 mbar is "very near the limit for human survival", according to Barometric pressures on Mt. Everest: new data and physiological significance, *John B. West 1999.*

**6:** At high oxygen partial pressure, organic materials will burn easily and rapidly, such as in the Apollo 1 disaster. However, the Apollo 7 mission used 345 mbar pure oxygen for 10 days, from the Apollo 7 Press Kit, *9/27/68*. Still, total flammability would likely be somewhat higher for hydrocarbon-derived materials then in Earth conditions, eg from Oxygen Partial Pressure and Oxygen Concentration Flammability: Can They Be Correlated?, *Harper et al. 2016.*

**7:** Simple models suggest that 20 mbar of oxygen would be sufficient to moderate Martian UV irradiance to present-day Earth values through the photochemical production of ozone, eg. The Ultraviolet Environment of Mars: Biological Implications Past, Present, and Future, *Cockell et al. 2000*. However, questions remain around the evolution of ozone in a Martian atmosphere, and 1-D photochemical modeling of the Martian atmosphere is needed to conclusively answer this question.

**8:** A 12 mbar (1200 pa) atmosphere would serve to block most solar energetic particles, including flares, as seen in figure 6 of What is the Radiation Impact of Extreme Solar Energetic Particle Events on Mars?, *Zhang et al. 2023.* using the radiation tolerances from Solar particle event storm shelter requirements for missions beyond low Earth orbit *Townsend et al. 2018.*

**9:** A 100 mbar atmosphere would provide a similar amount of shielding as 2.4 meters of regolith, as shown in figure 10 of Martian sub-surface ionising radiation: biosignatures and geology, *Dartnell et al. 2007.* This is similar to exposure levels on the ISS, and may require additional mitigation such as human genetic engineering to better tolerate ionizing radiation.

**10:** Ion Escape From Mars Through Time: An Extrapolation of Atmospheric Loss Based on 10 Years of Mars Express Measurements *Ramstad et al. 2018* shows a loss of 4.8 mbar over 3.3 billion years, even with higher solar flux in the distant past. Under these assumptions, 1% loss of our proposed atmosphere would take approximately 680 million years.

**11:** Figure on next page

**12:** This argument is explored in detail in Inventory of $CO_2$ available for terraforming Mars, *Jokosky & Edwards, 2018,* which concludes that "...there is not enough $CO_2$ remaining on Mars to provide significant greenhouse warming..."

**13:** On the secular retention of ground water and ice on Mars, *Dadile et al. 2020.* suggests that "If Mars' post-Noachian crustal $H_2O$ inventory was a few hundred meters GEL or more, then... groundwater likely exists globally on Mars today."

**14:** Demoed at the small scale & modeled at large scale in Biomaterials for organically generated habitats beyond Earth, *Wordsworth et al. 2025*

**15:** Published in Feasibility of keeping Mars warm with nanoparticles, *Ansari et al. 2024*. Will be a topic of further discussion at the workshop.

**16:** Microbial growth has been observed under simulated Martian atmospheric conditions, Growth of Carnobacterium spp. from permafrost under low pressure, temperature, and anoxic atmosphere has implications for Earth microbes on Mars, *Nicholson et al. 2012*. For plants, modeling has shown that plants should be capable of growing in 50 mbar atmospheres, Toward Martian Agriculture: Responses of Plants to Hypobaria, *Kenneth et al. 2002.*

**17:** As in Enabling Martian habitability with silica aerogel via the solid-state greenhouse effect, *Wordsworth et al. 2019.* using aerogels already shown to work on Mars in Development of a Thermal Control Architecture for the Mars Exploration Rovers , *Novak et al. 2003.*

**18:** This is an ongoing field of research, but there is promise for biological aerogels with high transparency and low thermal conductivity. Highly transparent silanized cellulose aerogels for boosting energy efficiency of glazing in buildings, *Abraham et al. 2023.*

**19:** As calculated with a 3-dimensional model in figure S13 of Atmospheric dynamics of first steps toward terraforming Mars, *Richardson et al. 2025.* 35 mg/m² x $1.44 \times 10^{14}$ m² (the surface area of Mars) = $5 \times 10^9$ kg.

**20:** Without an increased greenhouse effect from gaseous $CO_2$ this is evident from simple order-of-magnitude math. A 30° C temperature gain would increase the radiative flux by 80 W/m², and imply a total planetary power loss by $10^{16}$ watts. Assuming a warhead yield of about 300 kilotons or $10^{15}$ J, that would require detonating approximately 1.5 million nuclear weapons per day. This is not a firm estimate, but it should communicate the infeasibility of this approach.

**21:** In Radiative-convective model of warming Mars with artificial greenhouse gases, *Marinova et al. 2005.* estimated that it would take about 1 pascal of $C_2F_6$, $SF_6$ or a similar gas to warm Mars 30° C, which is approximately $7 \times 10^{15}$ moles of fluorine. However, according to The Role of Halogens During Fluid and Magmatic Processes on Mars, *Rampe et al. 2018,* fluorine is only present in the Martian regolith in the 15-90 parts per million range, which would require harvesting all of the fluorine in the top 40 meters of the planetary crust to attain sufficient quantities.

**11: Log-scaled version of elemental distribution plot, with numbers in teratonnes.** (1 teratonne = $10^{15}$ kg.)

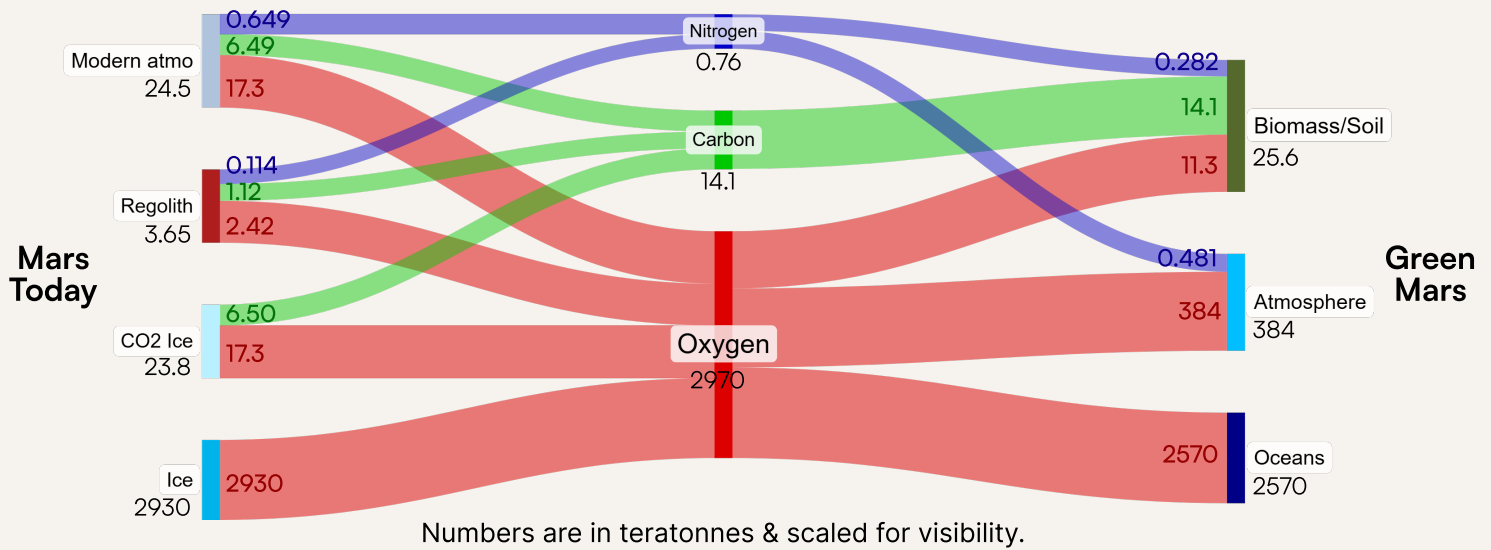

Numbers are in teratonnes & scaled for visibility.

Citations for the above plot:

**Atmospheric Data:** Mars Fact Sheet, *Dr. David R. Williams, 2025.*

**Water Ice Cap Data:** Comparison of the North and South Polar Caps of Mars: New Observations from MOLA Data and Discussion of Some Outstanding Questions, *Fishbaugh et al. 2001.* & Subsurface Radar Sounding of the South Polar Layered Deposits of Mars, *Pluat et al 2007.*

**CO₂ Ice data:** Flexure of the Lithosphere Beneath the North Polar Cap of Mars: Implications for Ice Composition and Heat Flow, *Broquet et al. 2020.*

**Regolith quantity** is modeled as a planet-wide layer 4 meters thick and a mass of $8.08 \times 10^{17}$ kg, according to Rocky ejecta craters as a proxy for the evolution of regolith on Mars, *Rajšić et al 2023.* & Exploring Regolith Depth and Cycling on Mars, *Fassett et al. 2017.*

**Regolith nitrogen & oxygen composition:** The nitrate/(per)chlorate relationship on Mars, *Stern et al. 2017.*

**Regolith carbon composition:** Combined carbonate and adsorbed carbon dioxide from Carbonates identified by the Curiosity rover indicate a carbon cycle operated on ancient Mars, *Tutolo et al. 2025* & Obliquity-Driven CO₂ Exchange Between Mars' Atmosphere, Regolith, and Polar Cap, *Buhler et al. 2021.*

**Biomass ratio:** Is assumed to be a combination of dead woody biomass & live microbial, from Elemental Composition (C, N, P) and Cell Volume of Exponentially Growing and Nutrient-Limited Bacterioplankton, *Vrede et al. 2002.* & Physicochemical Properties of Forest Wood Biomass for Bioenergy Application: A Review *Bianchini et al. 2025.*

**22:** A 150 mbar planetary atmosphere of oxygen requires $3.6 \times 10^{19}$ moles of oxygen. According to Ecology: Theories and Applications, *Peter Stiling. 1996* & The Ecology of Plants, *Gurevitch et al. 2002.* the productivity of boreal forest environments is 800 grams biomass/m²/year. Making one mole of CH₂O biomass produces roughly one mole of oxygen, so a reasonable oxygen production rate for a Green Mars is likely around $5 \times 10^{15}$ moles/year, or and it would require approximately 7500 years to generate a 150 mbar atmosphere. However, under a 20 meter dome only 100 moles/m² are needed for 150 mbar, which can be produced biologically in 3 years assuming similar levels of productivity.

**23:** Microbial growth with "100% Mars resources" using simulated Mars resources from regolith, water ice and atmosphere is shown in Our first target microbe for Mars, and how we chose it, *Pioneer Labs Reports. 2025,* with greater details on inorganic nutrients in A Biologist's Guide to Mars Dirt, *Pioneer Labs Reports 2024.*

**24:** Microbial production of aerogels is an early field, but shows promise as a tool to synthesize transparent insulating materials, Aerogels based on Bacterial Nanocellulose and their Applications, *Panahi-Sarmad et al. 2024.* & Synthesis of Acetobacter xylinum Bacterial Cellulose Aerogels and Their Effect on the Selected Properties, *Sozcu et al. 2025.* Additionally, bioplastic or biocement are both useful biological building materials.

**25:** Ice covered lakes are naturally buffered from UV (water is very UV absorptive) and from large temperature swings (water has high heat capacity, and ice is itself insulating), making ice covered lakes some of the most extreme inhabited environments on Earth. See Antarctic environments as models of planetary habitats, *McKay et al 2017*

**26:** Soil formation, or pedogenesis, involves biogeochemical processes that convert rock into soil through weathering, as described in Microbial Weathering of Minerals and Rocks in Natural Environments, *Samuels et al. 2020.* This is an important early stage of ecological succession.

# Appendix continued

**27:** Existence of endemic Martian life would create new possibilities for symbiotic ecosystems or organisms, Engineering Microbial Symbiosis for Mars Habitability, Correll and Worden 2025.

**28:** The Search for Life Science Analysis Group is intended to refine the recommendations for a Mars-lander to search for life in the 2023-2032 Plantary Science Decadal survey with four primary goals: 1. The search for direct evidence of extant life, 2. Assessment of near surface habitability, 3. Determining the history of processes and palaeoenvironments recorded in mid-latitude ice deposits, 4. Determining processes that preserve/modify/destroy ice deposits.

For more information, see Mars Life Explorer, *Amy Williams and Brian Muirhead, 2021.*

**29:** As reported in Italy Signs Agreement with SpaceX for Starship Mars Mission, *Andrew Parsonson, 2025.* which will send a plant growth experiment, a radiation sensor, and a meteorological monitoring station to collect data on the voyage to and surface of Mars.

**30:** The International Mars Ice Mapper (I-MIM) is a currently-paused mission to quantify the extent and volume of water ice, overbuden, and volatiles in non-polar regions within 5-10 meters of the surface to enable access by crewed missions The International Mars Ice Mapper (I-MIM) Mission: A Concept Advancing Climatology, Geology, Habitability, and Science for Human Exploration, *Ammanito et al. 2024.*

**31:** The Mars Telecommunications Orbiter, now being planned for 2028 by Blue Origin, has space for 1000 kg of payload to Mars orbit, which may include a climate module. Mars Telecommunications Orbiter, *Blue Origin 2025.*

**32:** As proposed in How to terraform Mars for $10B in 10 years, *Casey Handmer, 2022.* Note that the headline cost is for heating Mars 2 °C (~the amount Earth has been heated), not 30 °C (the amount likely needed for terraforming Mars).

**33:** In Plants grown in Apollo lunar regolith present stress-associated transcriptomes that inform prospects for lunar exploration, *Paul et al. 2022,* plant growth was observed with the addition of nutrient solution containing all necessary macro and nutrient salts, as well as sucrose and plant growth hormones.

## Present-day Mars statistics

**Handy reference guide for your order-of-magnitude calculations.** Compiled by Edwin Kite and duplicated from the 2024 workshop.

Planetary Characteristics: Top-of-atmosphere insolation 150 W/m² ($2 \times 10^{16}$ W). 6 mbar atmospheric pressure (170 kg/m²); >10 mbar at low elevation. 95% $CO_2$, 3% $N_2$, 2% Ar, 0.1% $O_2$, 0.1% CO (by volume)[1]; on average 10 precipitable μm of $H_2O$ vapor[2]. Average surface temperature -65 °C. Day-night temperature range at equator ~100 °C[3]. Surface UV-C&B (200-315 nm) flux 10× Earth[4]. Surface cosmic radiation (galactic+solar) 0.7 mSv/day[5]. Gravitational acceleration 38% of Earth. Rotation rate, obliquity, and land surface area: all same as Earth (±5%).

$H_2O$ ice is present at < 1m depth over one-third of Mars. Confirmed $H_2O$-ice volume is $5 \times 10^6$ km³ (35 m global equivalent layer), mostly in the two $H_2O$-ice polar caps (>80° latitude)[6-7]. Buried ice at <1 m depth is almost ubiquitous polewards of 40° latitude, but limited/patchy elsewhere[8]. "Dry" areas have 2±1 wt% water- equivalent hydrogen (from OH, H, and $H_2O$ in minerals)[9]. Buried liquid $H_2O$ aquifers: Unconstrained, but if present would be >km deep[10].

Soil: salty, basaltic, rich in P but poor in N and C. Details[11] (n=90, 1σ) - weight%: $SiO_2$=45.8±1.5, $TiO_2$=0.9±0.2, $Al_2O_3$=9.7±0.8, FeO=17.2±1.8, $Cr_2O_3$=0.4±0.1, MnO=0.4±0.04, MgO=8.2±0.9, CaO=6.6±0.45, $Na_2O$=2.7±0.5, $K_2O$=0.4±0.05, $P_2O_5$=0.9±0.1, $SO_3$=6.1±1.6, Cl=0.7±0.1. ppm by weight: Ni=474±135, Zn=284±91, Br=65±67. Volatiles include: oxychlorine phases, nitrates, organic C, sulfates[12]. Potential electron acceptors include: sulfate and hematite/magnetite in sedimentary rocks[13] (total volume $1.5 \times 10^6$ km³) and in soil.

Volcanically dormant: no hot springs today.

No aqueous-fluid surface flow today. Activity monitored from orbit is apparently not driven by aqueous fluids[14]. Big rivers+lakes dried up >$10^9$ yr ago. Theoretically, climates warm enough for liquid water flows on steep slopes might have occurred as recently as $3 \times 10^5$ yr ago[15-16].

Sources:
[1] Rafkin et al, chapter in "Comparative climatology of terrestrial planets" (2014), edited by Mackwell et al. [2] Montmessin et al., chapter in "The atmosphere and climate of Mars" (2017), edited by Haberle et al. [3] Martinez et al. J. Geophys. Research 2021. [4] Cockell et al. Icarus 2000. [5] Ehresmann et al. Icarus 2023. [6] Carr & Head Geophys. Research Letters 2015. [7] Levy et al., J. Geophys. Research 2014. [8] swim.psi.edu [9] Mitrofanov et al. J. Geophys. Research 2022. [10] Grimm et al. J. Geophys. Research 2017. [11] O'Connell-Cooper et al. J. Geophys. Research 2017. [12] Sutter et al., chapter in "Volatiles in the Martian crust" (2019), edited by Filibexrto & Schwenzer. [13] Rampe et al. Geochemistry 2020. [14] Dundas et al. J. Geophys. Research 2021. [15] Madeleine et al. Geophys. Research Letters 2014. [16] Dickson et al. Science 2023.